# Active Noise Reduction in Si/SiGe Gated Quantum Dots

[1][#]Rajat Bharadwaj, [1][#]Parvathy Gireesan, [1]Harikrishnan Sundaresan, [2,3]Chithra H Sharma, [1]Lucky Donald L Kynshi, [1]Prasad Muragesh, [4]D. Bougeard, and [1]Madhu Thalakulam[*]

[1]School of Physics, Indian Institute of Science Education and Research Thiruvananthapuram, Kerala 695551, India
[2]Christian-Albrechts-Universität zu Kiel, Institut für Experimentelle und Angewandte Physik, 24098 Kiel, Germany
[3]Center for Hybrid Nanostructures, Universität Hamburg, Luruper Chaussee 149, 22761 Hamburg 22761, Germany
[4]Fakultät für Physik, Universität Regensburg, Regensburg 93040, Germany

## Abstract

Solid-state quantum technologies such as quantum dot qubits and quantum electrical metrology circuits rely on quantum phenomena at ultra-low energies, making them highly sensitive to various forms of environmental noise. Conventional passive filtering schemes can reduce high-frequency noise but are often ineffective against low-frequency interference, like powerline or instrument-induced. Extending such filters to lower frequencies causes practical issues such as longer stabilization times, slower system response, and increased Johnson noise, which impede low-frequency transport measurements. To address these limitations, we propose and experimentally demonstrate a generalized active noise cancellation scheme for quantum devices operating at sub-Kelvin temperatures. Our approach compensates periodic environmental interference by dynamically injecting a phase-coherent anti-noise signal directly into the device. We employ an automated feedback protocol featuring beat-frequency reduction and adaptive phase-amplitude tuning, enabling real-time compensation without any manual intervention. Unlike post-processing or passive filtering, this method suppresses noise at the device level without introducing additional time constants. We implement the scheme on a gate-defined Si/SiGe quantum dot subject to strong 50 Hz powerline interference and validate its effectiveness through acquiring Coulomb Blockade Oscillations and Coulomb diamond plots. The technique achieves substantial suppression of both the targeted interference and the overall noise floor, thereby stabilizing transport characteristics and enhancing device fidelity. While demonstrated on a quantum dot, the proposed framework is broadly applicable to a wide class of solid-state quantum devices where deterministic noise presents a critical bottleneck. Our results establish active anti-noise injection as a versatile strategy for advancing noise-resilient quantum measurement platforms.

[#] Equal contribution
[*] madhu@iisertvm.ac.in

Quantum technologies such as solid-state qubits, quantum electrical sensing, and metrology circuits harness and harvest various quantum phenomena, whose energy scales are in a few tens to a few hundreds of micro-electron volts. These extreme low energy scales warrant their operation in sub-Kelvin, ultra-low noise environments to minimize decoherence and dephasing due to classical noise sources. Among solid-state quantum devices, gate-defined semiconductor nanostructures, such as gated quantum dot systems, are highly sought-after platforms for investigating fundamental physics and developing scalable quantum technologies[1–3]. However, their extreme sensitivity to the local electrostatic environment makes them particularly vulnerable to noise[4–7], which can arise from trapped charges, dielectric fluctuations, or interference from external sources. In mesoscopic systems such as quantum dots, even a small perturbation in the charge environment can lead to fluctuations in conductance, dephasing, and instabilities in device operation. Noises in the higher frequency ranges are filtered out using low-pass filters installed at various stages of the measurement system, such as the cryostat, just before reaching the device electrodes[8]. However, low-frequency unwanted signals, which could be a result of powerline interference, instrument clocks, periodic vibrations, etc., are non-trivial to filter. Usage of a low-pass filter operating in the lower frequency regimes, a few to a few tens of Hertz, can cause measurement delays and prolonged stabilization time during the experiment[9]. This will also forbid one from conducting low-frequency transport measurements such as lock-in detection. Although one can use lock-in amplifiers to achieve a cleaner signal, it is not a solution to reduce the noise imparted onto the device. Filters on the output side or post-processing algorithms[10] may smooth the measurement data, but do not mitigate the noise at the device level. Existing efforts to mitigate noise in quantum devices have largely focused on both passive and active approaches. Passive techniques often involve the use of low-pass filters, such as LC (inductor-capacitor), RC (resistor-capacitor), or distributed copper powder filters[11,12] placed at the input side of the device. While these filters are effective at attenuating high-frequency noise, reducing their cutoff frequency to target low-frequency interferences require the use of resistors and capacitors with large values, which is often a challenge. Apart from causing measurement delays and stabilization issues, the use of resistors can also impart more Johnson noise. The use of large electrolytic capacitors in sensitive quantum measurement circuits is always a non-trivial problem. Active noise-suppressing strategies, such as compensating power supply-induced magnetic field fluctuations using auxiliary coils, have shown promise in trapped ion qubits[13,14], motivating a broader exploration of active approaches. One of our earlier works[15] attempted to mitigate low-frequency interference using a predictive input shift, but its efficiency was limited by beat distortion and computational delays, typically below 1 Hz.

Active noise cancellation, which suppresses unwanted signals by injecting phase-coherent anti-noise, offers a broadly applicable framework across diverse systems. In this context, we introduce a scheme suitable for systems that enables characterization of the noise profile and provides electrical access to the affected regions for the injection of anti-noise signals, thereby compensating for unwanted noise. While the method is applicable across all frequency ranges, it is particularly well-suited for low-frequency interferences, where passive filtering proves inadequate. We demonstrate this approach on a gate-defined quantum dot (QD) formed in a doped Si/SiGe heterostructure[7,16–27], subject to 50 Hz powerline interference, and show substantial noise suppression through comparative measurements of Coulomb Blockade Oscillations (CBOs) and Coulomb diamond plots[2,22,26,28], both with and without compensation. Furthermore, we examine how the effectiveness of the compensation varies across different operational regimes of the quantum dot.

The device used in this study is an array of six gate-defined quantum dots, fabricated on a doped Si/SiGe heterostructure[7,16,27] hosting a two-dimensional electron gas (2DEG), as shown in Fig. 1(a). The measured carrier concentration and mobility of the dopant-induced 2DEG in Hall bar geometry devices are $\sim 2.4 \times 10^{11} cm^{-2}$ and $\sim 48,000\ cm^2/Vs$, respectively. All measurements are performed with the device mounted on the 10 mK stage of a cryogen-free dilution refrigerator. In this work, we used only the top-left QD, defined by energizing the depletion-gates highlighted in red. The regular oscillations in dot current against the gate voltage, the Coulomb blockade oscillations (CBO), as shown in Fig. 1(b) is a signature of the formation of a well-defined quantum dot on our device[17]. The black trace represents the CBO measured with a source-drain bias of 3 mV using a 1 Hz low-pass output filter, while the red trace shows the same acquired with a 10 kHz acquisition bandwidth. We find that the CBO acquired using the 10 kHz bandwidth is noisy and jittery, highlighting the impact of noise transmitted through the sample from the room temperature. Fig. 1(c) shows the frequency spectrum of the noise obtained for the dot operating point represented by the black arrow in Fig. 1(b). We find that the dominant 50 Hz and its weaker harmonics contribute to the noise.

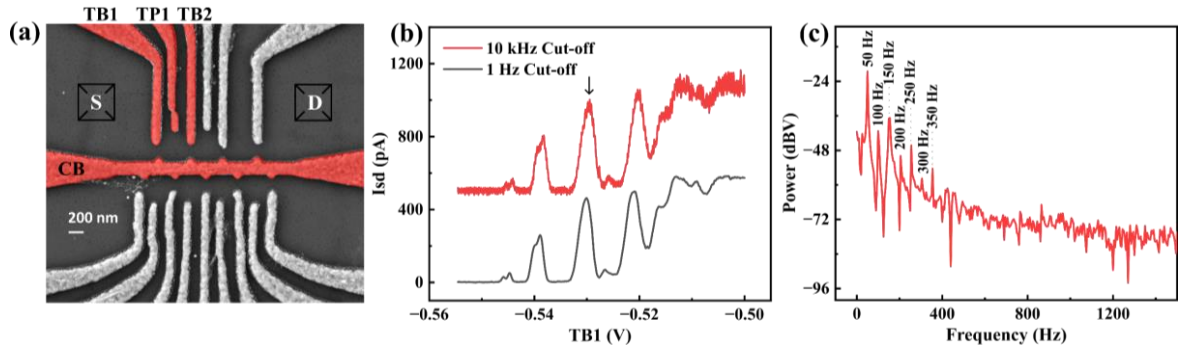

FIG. 1. (a) Micrograph of the Si/SiGe device similar to what is used in this study, featuring six quantum dots. This work focuses on the upper-left dot, defined by the central barrier gate (CB), top barrier gates (TB1 and TB2), and a plunger gate (TP1), highlighted in red. (b) Coulomb blockage oscillations as measured in the upper left dot with a low-pass filter placed at the output with cut-offs 1 Hz (black) and 10 kHz (red). (c) Noise spectrum measured at the peak indicated by the arrow in (b).

For minimising the effect of such a noise pattern, we make a scheme targeting the dominating frequency components, as summarized in Fig. 2. The instrumentation to implement the proposed procedure is given in the Supplementary Information-S1. The proposed procedure is outlined in the flow-chart shown in Fig. 2(a). First, the dominant frequency component of the noise, $f_k$, is identified from the major noise spectrum $\{f_1, f_2 \ldots, f_n\}$, acquired from the drain current, as shown in Fig. 2(b). Subsequently, a signal at the same nominal frequency is generated, denoted $f'_k$ referred to here as the anti-noise signal and injected into the device through the source. Ideally, $f'_k = f_k$. However, due to instrumental imperfections, a slight mismatch often exists between the intended output frequency $f_k$ and the actual output $f'_k$, which appears as a nearby peak in the frequency spectrum as shown in Fig. 2(c). The resultant signal in the device can then be represented with Eq. (1).

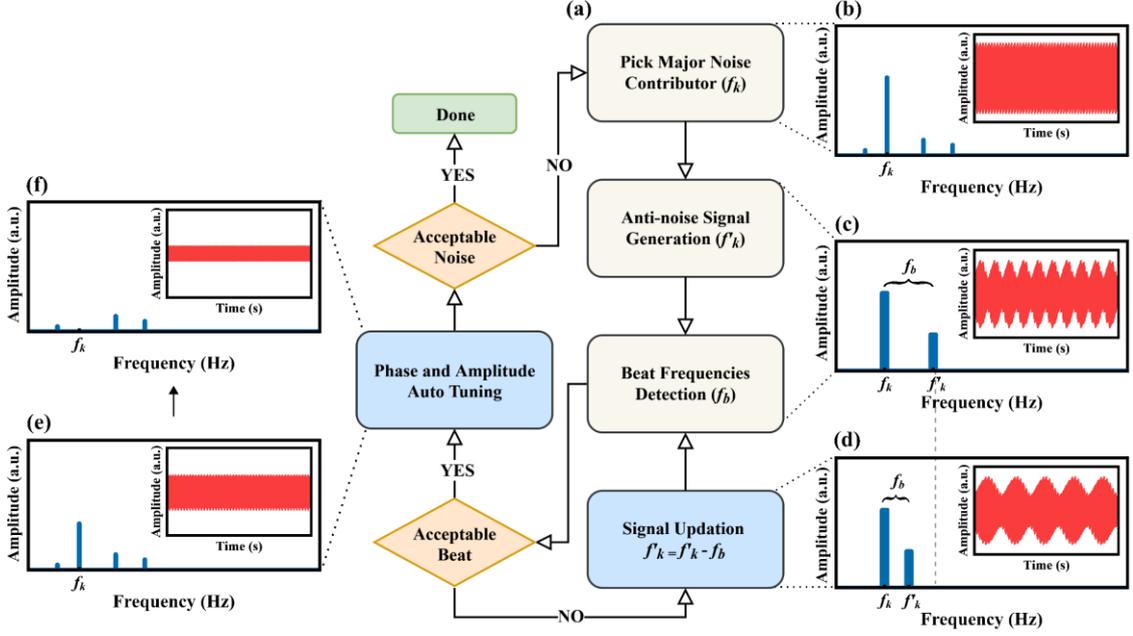

FIG. 2. (a) Proposed flow diagram of the procedure. (b) Shows the noise spectrum with time domain representation in the inset. (c) Mismatch between the synthesized signal $f'_k$ and actual noise component $f_k$. The inset shows the observed beat pattern having a period $1/f_b$. (d)The mismatch reduces after adjustment in $f'_k$ resulting in reduced $f_b$. The inset shows the increased beat duration. (e) Shows the nearly constant amplitude due to sufficiently low $f_b$. After auto-tuning the phase and amplitude, the compensation significantly reduces noise due to the component $f_k$ as shown in (f).

$$y(t) = A_1 \cos(2\pi f_k t + \phi_1) + A_2 \cos(2\pi f'_k t + \phi_2) \qquad (1)$$

This leads to the formation of a beating pattern in time-domain, with a frequency $f_b = |f_k - f'_k|$, as illustrated in the inset to Fig. 2(c). Minimizing this beat frequency is an important step in this compensation procedure. Even when the phase and amplitude are perfectly tuned, i.e., $A_1 = A_2 = A$ and $\phi_2 = \phi_1 + \pi$ (i.e., opposite phase), the resulting interference leads to the signal $y(t)$, as shown in Eq. (2).

$$y(t) = -2A\sin\left(2\pi \frac{f_b}{2} t\right) \sin\left(2\pi \frac{f_k + f'_k}{2} t + \phi_1\right) \qquad (2)$$

From Eq. (2) we find that the temporal window over which the compensation remains effective is $t = \frac{1}{6f_b}$. To increase the duration, the set frequency value is adjusted iteratively to bring $f'_k$ closer to $f_k$, as shown in Fig. 2(d). During this time window, the resulting amplitude of the carrier stays lesser than the amplitude of the existing noise. Once the beat frequency is minimized, the phase and amplitude of $f'_k$ require optimization. Phase mismatches arise due to lack of information of the noise phase, while amplitude discrepancies typically result from unequal attenuation paths in the circuit. To compensate for these effects, both phase and amplitude are adaptively tuned. The phase is adjusted first to be approximately $\pi$, out of phase with the noise at $f_k$, and the amplitude is then tuned to get the minimum resultant amplitude. This compensation is achieved using an adaptive tuning procedure illustrated in Supplementary Information-S2. In this exercise, the tuneable quantity $M$ (either phase or amplitude) is incrementally varied by a predefined step size $\Delta M$. After each step, the resulting change in noise power $\Delta P(f_k)$ is measured. If the power decreases, the same $\Delta M$ is applied in the next iteration. If no further improvement is observed, the direction of $\Delta M$ is reversed and its magnitude halved, ensuring a continued movement toward the optimum point. The process is repeated until the step size falls below a threshold $|\Delta M_{th}|$, indicating convergence. The impact of this tuning is illustrated in the transition from Fig. 2(e) to Fig. 2(f), where the targeted frequency component is finally fully compensated. Similarly, the optimal phase and amplitude are determined for each dominant component, $\{f_1, f_2, ... f_N\}$, the corresponding anti-noise must be synthesized as an arbitrary waveform combining all of these components. After this compensation is applied, the actual measurement can be performed.

Now we focus on the implementation of the noise compensation procedure on the QD device described in Fig. 1(a). The key components of the instrumentation used for the noise compensation procedure are shown in Supplementary Information-S1. We use a digital oscilloscope to characterize the noise on the CBO peak indicated by the black arrow in Fig. 1(b), and the noise spectrum obtained from conducting fast Fourier transform (FFT) is shown in Fig. 1(c). As a proof of concept, here, we target reducing the dominant 50 Hz component. The anti-noise signal is generated by the AWG and added to the sourcing side using a summing amplifier. Fig. 3(a) shows the resulting signal with a beat-frequency ~ 3 mHz (~ 329-second beat interval) in the drain current. We iteratively minimized this beat-frequency to obtain a beat interval of more than 1200 s, resulting in an effective compensation window of 200 s. Subsequently, the phase and amplitude of the compensation signal are programmatically auto-tuned to minimize the resultant noise power. The combined effect is illustrated in the time-domain plots in Fig. 3(b). The blue trace shows a significant reduction in the amplitude of the noise signal, at ~ 50 Hz, compared to the uncompensated red trace. The corresponding frequency-domain response before and after compensation is shown in red and blue in Fig. 3(c). The inset highlights the reduction in the ~ 50 Hz component, which undergoes a reduction of ~ 20 dBV. Compensation for other frequency components is not performed for quicker measurements, since their contribution is not substantial compared to the 50 Hz peak.

The magnitude of spectral noise reduction is further quantified in Supplementary Information S3(a), showing that the proposed approach not only suppresses the 50 Hz peak but also lowers the overall noise floor by stabilizing the electrochemical potential of the reservoir. For further confirmation, we calculate the total noise power after compensation using anti-noise signals with in-phase and out-of-phase configurations over the 0 to 1.5 kHz range, to evaluate their effect on white noise. The corresponding spectra and the total noise power comparison are provided in Supplementary Information S3(b) and S3(c), respectively. We find that the integrated noise power, in terms of $V^2$, decreases by an order of magnitude, from $5.24 \times 10^{-3}$ to $4.55 \times 10^{-4}$ (more than threefold reduction in the voltage noise) when anti-noise is applied with a 180° phase shift. In contrast, in-phase (0°) signal leads to an increase in noise power to $2.11 \times 10^{-2}$ due to constructive interference. Noise, especially within transport channels, can raise the effective electron temperature[29] and thereby degrade quantum coherence[30–32] and device fidelity. The observed reduction in overall noise power and background noise floor can lead to a lower effective electron temperature and improved coherence times.

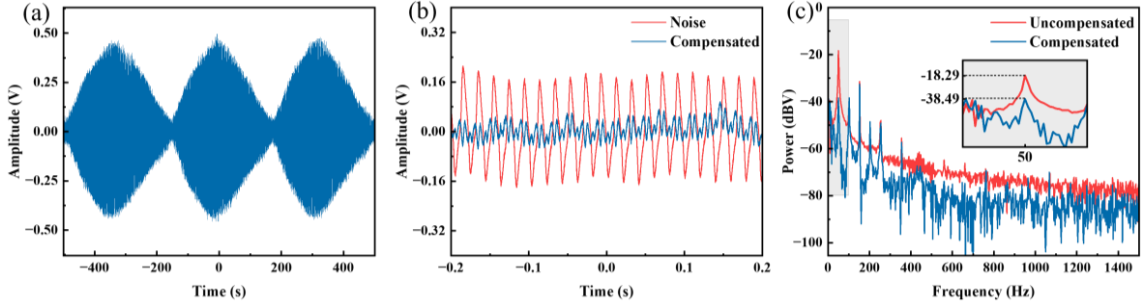

FIG. 3. (a) The beat signal as the consequence of the noise and anti-noise mismatch. (b) Time-domain signal showing the raw noise (red) and the signal after applying noise compensation (blue). Raw signal dominates with 50 Hz noise. (c) Noise spectrum before (red) and after (blue) compensation at the CBO peak indicated by the arrow in Fig. 1(b). The inset shows the marked grey area indicating the dominant contribution of this frequency to the overall system noise.

Now we explore the applicability of our technique to acquire Coulomb diamond plots, a touchstone characteristic of quantum dot formation, at broader acquisition bandwidths for the device shown in Fig. 1(a). Figure 4(a) shows the reference Coulomb diamond plot taken with a 1 Hz output filter. The data corresponding to the region enclosed by the white box in Fig. 4(a) is remeasured at a higher acquisition bandwidth of 1 kHz and is shown in the left panel of Fig. 4(b), while the same data acquired by applying the noise compensation technique is shown in the right. We perform phase auto-tuning after every 4 CBO sweeps which in total took 92.8 s (less than half of the compensation window) to speed up the compensated data acquisition. The overlayed white traces in Fig. 4(b) show the line-profiles representing the CBO along $V_{SD} = 0.760\ mV$, indicated by the dashed arrow in Fig. 4(a). Comparing the data in the left and right panels of Fig. 4(b), we learn that there is a clear reduction in the overall spurious signals in the data after the compensation procedure. We also find that the noise-compensated data acquired with a 1 kHz filter setting is comparable to that acquired with a 1 Hz acquisition bandwidth, underscoring the efficacy of the adapted procedure.

The effectiveness of the noise compensation appears to depend on the channel conductance since the amplitude of the anti-noise signal is determined from the noise in the drain current for a specific device operating point. Fig. 4(c) shows the noise-power observed at the pre-amplifier output versus the gate voltage, extracted from the CBO data. The compensating amplitude and phase applied for each sweep here are based on the noise spectrum observed on the first CBO peak indicated with an arrow in Fig. 4(b). Since the anti-noise is determined from the noise spectrum, its efficacy will also vary as we move to different conductance regimes. This is also evident from the post-compensated noise power, the blue trace in Fig. 4(c), which is lowest at the first CBO peak. While this may initially appear to be a limitation, in practice, in gated quantum dot experiments, the gate voltages are rarely swept over large ranges once the operating regime or the charge state is identified. In addition, the anti-noise amplitude can be adaptively tuned to lessen the effect due to the variations in conductance, if required.

Figure 4(d) further elaborates this behaviour. The traces show the ~ 50 Hz noise power against the phase difference between the signal and the anti-noise, observed at points labelled 1, 2, 3, and 4, along the CBO. Here, the amplitude of the anti-noise signal is calibrated based on the noise observed at point 1 and kept purposefully the same for the others. For 180° phase difference, we note that the noise power is minimum for the operating point labelled 1, where the compensation efficiency is maximum. The compensation becomes less effective as the conductance of the system deviates from the reference operating point, point 1 in this case, for the same anti-noise amplitude. This is also a touchstone proof that the compensation process is working as per the procedure outlined in Fig. 2.

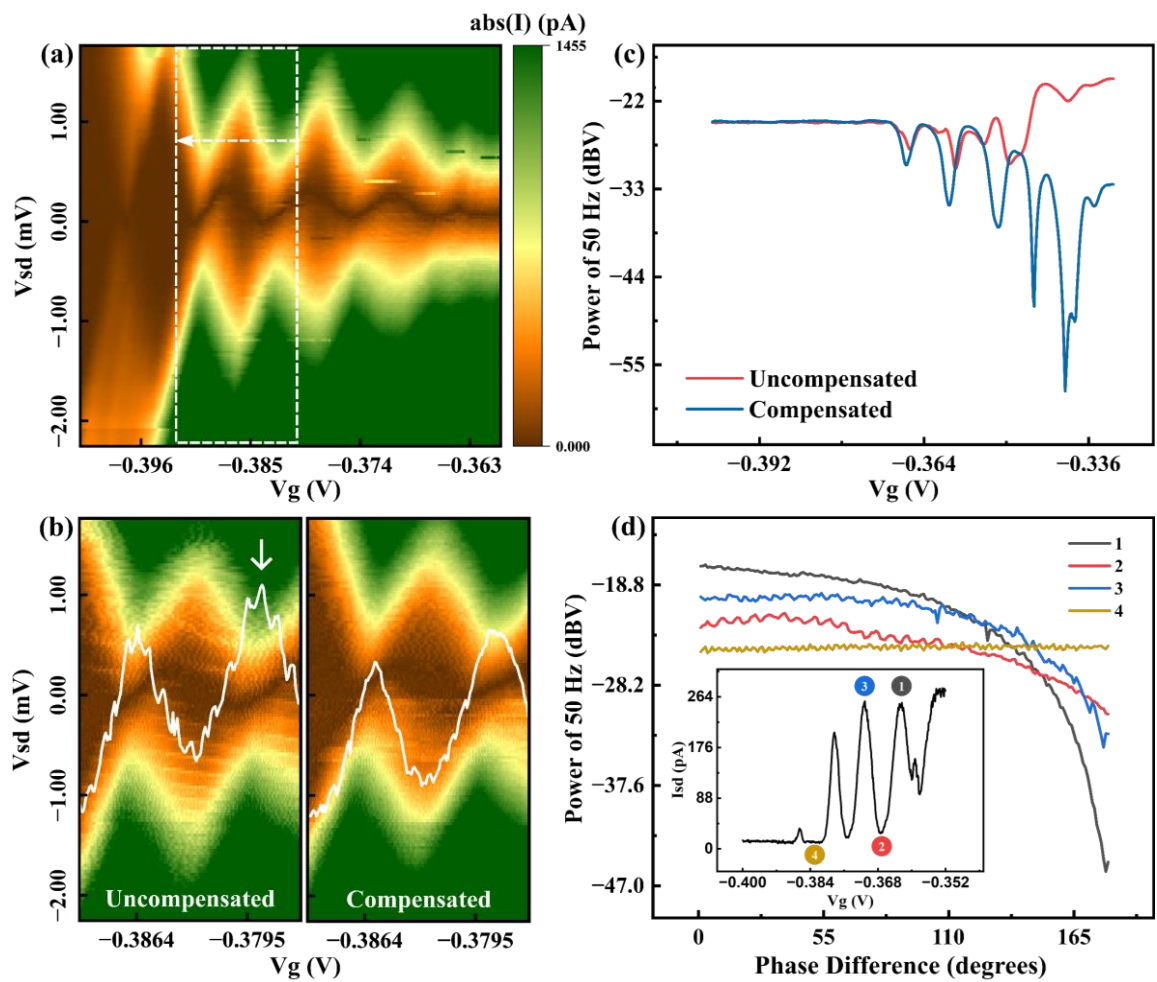

FIG. 4. (a) Reference Coulomb diamond plot measured with a 1 Hz output filter. (b) Comparison of uncompensated and compensated measurements within the dashed region of (a), using a 1 kHz output filter to make the effect of 50 Hz noise visible. The curve along the arrow is also plotted. (c) Shows the behaviour of 50 Hz power with (blue) and without (red) compensation along the CBO peaks. (d) Power of 50Hz frequency at output vs phase difference between noise and anti-noise signal at various CBO points, keeping a constant anti-noise amplitude auto-tuned for peak 1.

In this work, we presented an active noise cancellation scheme for quantum devices to suppress environmental interferences, such as ground loops and electronic jitters. The method involves the injection of an anti-noise signal, resulting in the cancellation of the noise signal on the device. Unlike conventional filtering approaches, which are often ineffective at low frequencies due to delayed system response and prolonged stabilization times, our technique is particularly effective at compensating low-frequency noise without compromising measurement speed or stability. The cancellation process is fully automated, leveraging beat frequency suppression and phase-amplitude auto-tuning for real-time adaptation. We experimentally validated the approach on a Si/SiGe quantum dot system, operated in a noisy environment, focusing on the dominant ~50 Hz environmental interference measured through the drain. Since the gate-lines are heavily filtered to preserve qubit coherence, and the high-frequency signals are applied through bias-tees with 100s of kHz cut-off, we inject the anti-noise through the source line. Measurements of CBOs and Coulomb diamonds at 10 mK revealed significant noise suppression. We find that not only the 50 Hz component, but also the overall noise floor showed a reduction. The observed reduction also may point to the fact that the compensation as envisioned may lead to lower electron temperatures, especially in cases where the noise is coupled to the bias line, a work which needs to be addressed in the future. The compensation leads to reduced fluctuations in the electrochemical potential of the reservoir, resulting in reduced charge noise, and is expected to improve coherence times[30,31]. Although demonstrated on a quantum dot device, the proposed scheme can be extended to other systems where deterministic noise can be independently characterized and a suitable anti-noise injection pathway is available. This is particularly valuable in quantum devices, where deterministic noise can otherwise distort output signals and perturb intrinsic device characteristics.

**Acknowledgements:** MT acknowledges funding support from the National Quantum Mission, an initiative of the Department of Science and Technology, Ministry of Science and Technology, India.

# Active Noise Reduction in Si/SiGe Gated Quantum Dots

[1#]Rajat Bharadwaj, [1#]Parvathy Gireesan, [1]Harikrishnan Sundaresan, [2,3]Chithra H Sharma, [1]Lucky Donald L Kynshi, [1]Prasad Muragesh, [4]D. Bougeard, and [1]Madhu Thalakulam[*]

[1]School of Physics, Indian Institute of Science Education and Research Thiruvananthapuram, Kerala 695551, India
[2]Christian-Albrechts-Universität zu Kiel, Institut für Experimentelle und Angewandte Physik, 24098 Kiel, Germany
[3]Center for Hybrid Nanostructures, Universität Hamburg, Luruper Chaussee 149, 22761 Hamburg 22761, Germany
[4]Fakultät für Physik, Universität Regensburg, Regensburg 93040, Germany

## Supplementary Material

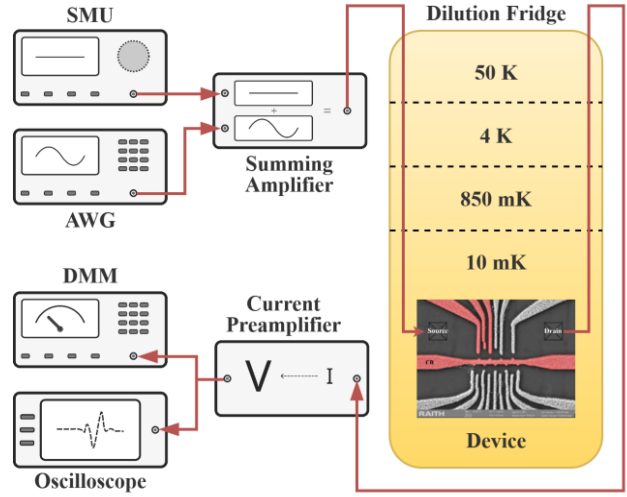

**S1.** Experimental setup for noise compensation. The DAC and AWG are connected to a summing amplifier, which is subsequently connected to the source electrode of the quantum dot, located at the 10 mK stage of a dilution refrigerator. The drain is connected to a current preamplifier that amplifies and converts the current into a voltage signal. This signal is then simultaneously routed to a digital multimeter (DMM) and an oscilloscope for measurement and analysis.

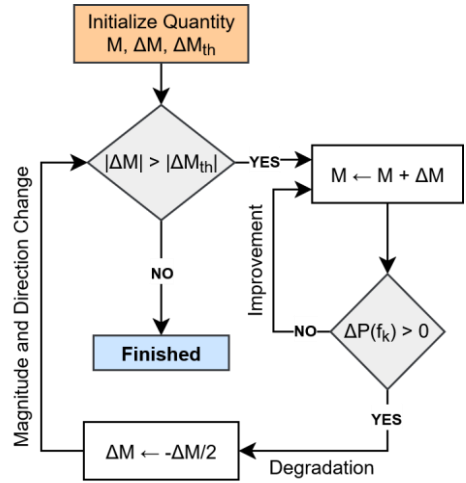

**S2.** Autotuning flow for a quantity $M$. $\Delta M$ and $\Delta M_{th}$ represent the update step size and its tolerance threshold, respectively. $\Delta P(f_k)$ represents the difference between improved power and previous power for the targeted frequency component $f_k$.

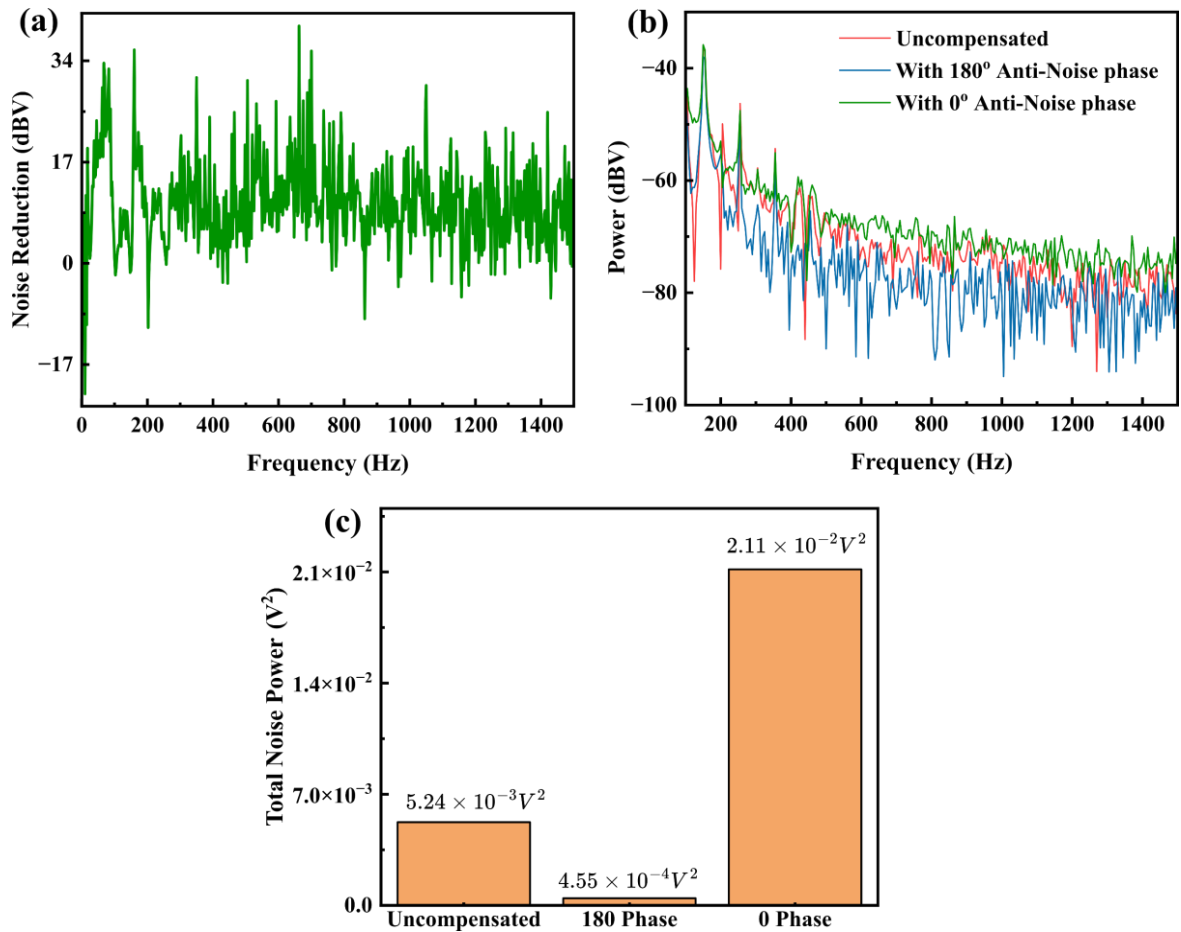

**S3.** (a) The difference between the uncompensated and compensated (with 180° phase difference) spectrums shown in (b). Having almost all the points above zero shows that the compensation is effective throughout the spectrum. (b) Noise spectra over the 0–1.5 kHz bandwidth for various phases of injected anti-noise relative to the existing 50 Hz signal. The blue trace, corresponding to anti-noise with a 180° phase shift, shows a clear reduction in overall noise, including white noise. In contrast, the green trace shows an amplification of the noise spectrum compared to the uncompensated case, shown in red. (c) Corresponding total noise power.